# Anharmonic thermodynamics redefines metastability and parent phases in ferroelectric HfO$_2$

Yiheng Shen, Chang Liu, Wei Xie,* and Wei Ren*

**Author Information**

Corresponding authors

Wei Xie – Materials Genome Institute, Shanghai University, Shanghai 200444, China; Email: xiewei@xielab.org

Wei Ren – Materials Genome Institute, State Key Laboratory of Advanced Refractories, Institute for Quantum Science and Technology, Physics Department, Shanghai University, Shanghai 200444, China; Email: renwei@shu.edu.cn

Authors

Yiheng Shen – Materials Genome Institute, Shanghai University, Shanghai 200444, China

Chang Liu – Materials Genome Institute, State Key Laboratory of Advanced Refractories, Institute for Quantum Science and Technology, Physics Department, Shanghai University, Shanghai 200444, China

**Abstract**

Hafnia ($HfO_2$) is a silicon-compatible dielectric material, yet stabilizing its desired but metastable ferroelectric phase remains challenging. Phase stability predictions by density functional theory (DFT) have provided crucial guidance, but most simulations neglected or only treated finite temperature effects with (quasi-)harmonic approximation due to high computational cost of DFT. Here, we develop a machine learning force field and perform thermodynamic calculations for $HfO_2$ using self-consistent phonon theory to address growing evidence of anharmonicity. Our results reveal that the ferroelectric orthorhombic phase oIII exhibits metastability below $0.1k_BT$ under most conditions within the simulated regime of temperature and pressure (600 K ≤ $T$ ≤ 1500 K and 0 ≤ $p$ ≤ 7.5 GPa), contradicting previous harmonic predictions of metastability above 1500 K at ambient pressure. We further report evidence for temperature- and pressure-dependent ferroelectric parent phase despite efforts to identify a universal one. This study highlights the importance of anharmonicity and provides an effective approach for its treatment in the design of $HfO_2$-based ferroelectrics.

**Main Text**

The discovery of ferroelectricity in $HfO_2$[1,2] which is readily silicon-compatible has spurred intensive researches[3-6] towards developing $HfO_2$ based electronic devices like ferroelectric random-access memories and ferroelectric field effect transistors. Nevertheless, practical applications of $HfO_2$ are hindered by the metastable nature of the desired ferroelectric phase,[7] which directly contributes to the wake-up[8] and fatigue effects[9] responsible for device performance degradation. Many recent efforts[10-14] are devoted to obtain kinetically stable ferroelectric $HfO_2$, in thin films or in the bulk, by engineering doping, stress, pressure, temperature, defects, surface/interface, electrode and electrical field etc. for better synthesis, purification and stabilization of the ferroelectric phase.

Such engineering development can be facilitated with the guidance of the pressure-temperature (*p-T*) phase diagram of $HfO_2$ first published in 2001.[7] At ambient temperature and pressure, $HfO_2$ exists in the monoclinic phase (*m*) with the *P*2$_1$/*c* space group. It undergoes a phase transition to the tetragonal phase (*t*) with the *P*4$_2$/*nmc* space group at approximately 1700°C, and further to the cubic phase (*c*) with the $Fm\bar{3}m$ space group at approximately 2500°C. Under compression at room temperature, $HfO_2$ transforms to two orthorhombic phases (oI and oII) with the *Pbca* and *Pnma* space groups at pressures of approximately 4 GPa and 14 GPa, respectively, with the oI phase being antipolar. The well recognized ferroelectric orthorhombic phase (oIII) with the *Pca*2$_1$ space group is a metastable phase and, therefore, not shown in the phase diagram. The potential energy surface of $HfO_2$ is, however, much more complex than shown therein. So far, HfO$_2$ has been suggested to crystallize in 16 stable or metastable polymorphs, and many of them having been confirmed in experiment. Among the latest experimental findings, the orthogonal nonpolar oI* with the *Pbca* space group[15] could play a key role the in ferroelectric switching pathway of oIII-HfO$_2$.[16,17] Additionally, other metastable ferroelectric phase has been observed in the hafnia-zirconia (HZO) systems, which might be the rhombohedral phase (r) with the *R3m* space group[10,18] or the orthorhombic phase (oIV) with the *Pmn2$_1$* space group.[11]

Meanwhile, previous simulations typically describe the zero temperature behavior of the HfO$_2$ lattices using density functional theory (DFT),[19] explaining the experimentally identified essence of oxygen vacancy[20] and epitaxial strain[21] with the static potential energy surface,[12,22] energy levels[11,23] and transition barriers.[24,25] However, the inherent inconsistence in the simulated temperature causes systematic error when describing finite-temperature properties with zero-temperature approaches, e.g. vibrational fingerprint *via* zero-temperature lattice dynamics calculations,[26] which neglects the temperature effects represented by phonon renormalization[27-29] and thermal expansion[30] and essentially leads to biased thermodynamic quantities.[23] Alternatively, finite-temperature molecular dynamics simulation provides phenomenal statistics over the transition kinetics, typically facilitated by defects,[31] lacking thermodynamic insights. Recently, first-principles thermodynamic calculation with consistent simulated temperature has been proposed,[32] though its heavy computations restrict it to systems with one degree of freedom (DOF), e.g., the cubic phase of HfO$_2$,[33] whereas first-principles thermodynamic calculations over multiple phases are not reported to our best knowledge.

In this work, stimulated by the advances in equivariant machine learning force fields (MLFF)[34] with DFT accuracy, we performed MLFF-accelerated thermodynamic calculations on the ferroelectric oIII-HfO$_2$ and its competing paraelectric and antipolar phases of primary concern in experiments, effectively evaluating the anisotropic thermal expansion and thermodynamic phase equilibrium.

**Equivariant MLFF as model for multiple HfO$_2$ phases.** We used the Monte Carlo modified rattle procedure as implemented in the hiPhive package[35] to generate a total of 240 rattled supercells of HfO$_2$, which are equally distributed to the pristine and compressed/stretched lattices (i.e., equiaxial strains of 0% and ±2%) of *c*-HfO$_2$, *t*-HfO$_2$, *m*-HfO$_2$, oI-HfO$_2$, oII-HfO$_2$, oIII-HfO$_2$, oIV-HfO$_2$ and *r*-HfO$_2$. Each rattled supercell is congruent with the 2×2×2 supercell of the cubic-phase unit cell containing 96 atoms. Check Note S1 in the Supporting Information (SI) for more details. 10% of the rattled structures, equally distributed among phases and strains, are selected as the test set, whereas the rest are used as the training set. A MLFF with the MACE architecture[34] featuring equivariant graph neuron network was trained upon the calculated energy and interatomic forces in the training set, yielding $R^2$ of 0.9998 (0.9989) and mean absolute error (MAE) of 38.2 (74.6) meV/Å for forces on the training (test) set, as summarized in Figure 1a (1b). The phase-resolved statistics in Figure S2-S3 in the SI further confirms the performance of this MACE MLFF in each concerned HfO$_2$ phase. We further validate the MACE MLFF by calculating the phonon spectra of HfO$_2$ with the finite displacement approach implemented in the Phonopy package,[36] as shown in

Figure 1c. The perfect consistence in the phonon spectra from VASP and MACE evidences the reliable description of the nuances in the interatomic forces by the MACE MLFF.

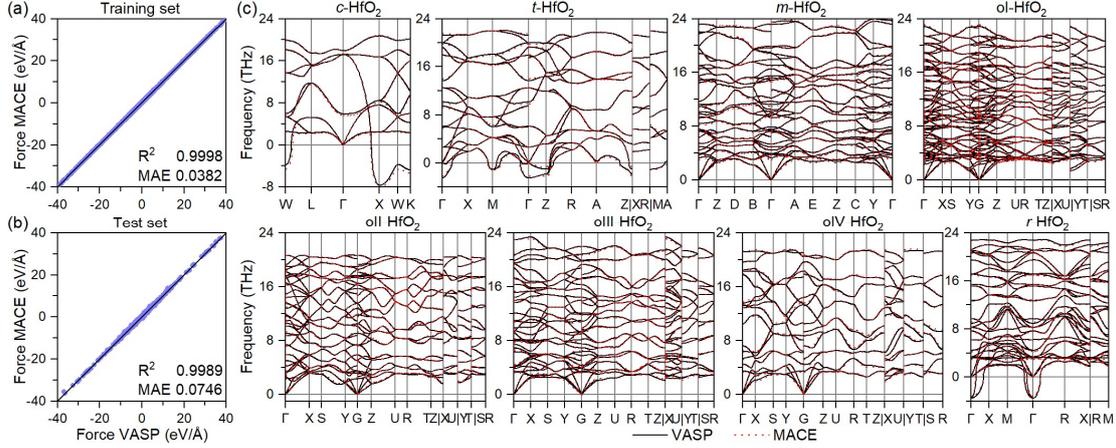

**Figure 1.** Parity plot for the interatomic forces in rattled HfO$_2$ supercells on (a) training and (b) test set. (c) Harmonic phonon spectrum of eight solid phases of HfO$_2$ calculated by VASP and the MACE MLFF.

**Simulating *p-T* phase diagram *via* thermodynamic calculations.** By neglecting the temperature dependence of the apparent atomic coordinates in a fixed lattice, we obtained Helmholtz free energy $F(\mathbf{A}, T)$ from the self-consistent phonon (SCP) calculations on atomic configurations optimized by DFT, where $\mathbf{A} = [\mathbf{a}, \mathbf{b}, \mathbf{c}]^T$ is the matrix formed by lattice vectors. We further obtain Gibbs free energy $G(T, p)$ and equilibrium lattice $\mathbf{A}_{eq}(T, p)$ by minimizing $F^* = F(\mathbf{A}, T) + pV$ over $\mathbf{A}$ at each given pressure $p$, where $V = \det(\mathbf{A})$ is the unit-cell volume. In practice, a weighted third-order polynomial fit of the $F^*$ against $\mathbf{A}$ was performed to facilitate the minimization, with an iterative optimization over the non-negative hyperparameter $x$ in exponential weight function $w = \exp(-xF^*)$ till the weighted MAE is lower than 1 meV/f.u., such that the error from the biased sampling of lattice parameters is addressed.[37]

Facilitated by the MLFF, we performed SCP calculations on pristine and strained lattices of *c*-, *t*-, *m*-, oI-, oIII- and oI\*-HfO$_2$ (see Figure 2 for a themed collection of SCP spectra and Figure S4-S9 in the SI for full results), which are commonly observed or expected in devices.[8,14] The non-analytical correction (NAC) was included in the SCP iterations, as detailed in Note S2 in the SI. For *t*-, *m*-, oI-, oIII- and oI\*-HfO$_2$ with multiple DOFs in lattice parameters, a mesh weaved by these DOFs are scanned to account for their coupling during anisotropic thermal expansion. In general, the renormalized phonon frequency in a given lattice increases with temperature and decreases upon expansion. The acoustic modes and the highest optical modes are almost inert with temperature, whereas modes between them significantly evolves with temperature, as represented by the wide span of the rainbow-colored bands in Figure 2a, evidencing profound anharmonicity in the solid HfO$_2$ phases and highlighting the essence of phonon renormalization in the thermodynamic calculations on these systems. Notably, *t*-HfO$_2$ gains Γ-, Z-, M- and X-point dynamical instability upon minor expansion (second panel of Figure 2a and Figure S5 in the SI), supporting the proper-ferroelectric nature of HfO$_2$ mediated by multiple plausible intermediate phases.[24,38,39] Based on the renormalized phonon spectra, we evaluated phase-resolved Gibbs free energy and equilibrium lattice parameters as a function of temperature and pressure. The latter is shown in Figure S10 in the SI. To highlight the necessity to include anharmonicity in our calculations, we display the anharmonicity measure[40] of each simulated HfO$_2$ phase in its equilibrium lattice, as shown in Figure 3a, with the overall and mode-resolved anharmonicity measures respectively defined as

$$\sigma^{A}(T)=\sqrt{\sum_{I}\left\langle\left(\mathbf{F}_{I}^{A}\right)^{2}\right\rangle_{T}\bigg/\sum_{I}\left\langle\left(\mathbf{F}_{I}\right)^{2}\right\rangle_{T}}$$

and

$$\sigma_{S}^{A}(T)=\sqrt{\sum_{s\in S}\left\langle\left(F_{s}^{A}\right)^{2}\right\rangle_{T}\bigg/\sum_{s\in S}\left\langle\left(F_{s}\right)^{2}\right\rangle_{T}},$$

where $F_s = \sum_I M_I^{-1/2} \mathbf{e}_{sI} \mathbf{F}_I$ is the atomic force $\mathbf{F}_I$ on atom $I$ projected on the phonon eigenvector $\mathbf{e}_{sI}$ of mode $s$, $F_s^A$ ($\mathbf{F}_I^A$) is the anharmonic contribution to $F_s$ ($\mathbf{F}_I$) beyond the converged effective harmonic model, $S$ is a collection of degenerate phonon modes, and $\langle \cdot \rangle_T$ is the ensemble average. Despite the common belief that dynamically stable m-, oI-, and oIII-$HfO_2$ can be effectively described in a harmonic model, they all possess moderate anharmonicity measures around 0.25-0.45 in the simulated temperature range, higher than that in crystalline silicon with strong harmonicity.[40] Mode-resolved analysis (Figure 3b) shows that the anharmonicity in $HfO_2$ is equally contributed by all phonons regardless of the temperature dependence of their phonon frequencies, indicating that the SCP-based model for thermodynamics is a major overhaul rather than minor corrections to the harmonic model.

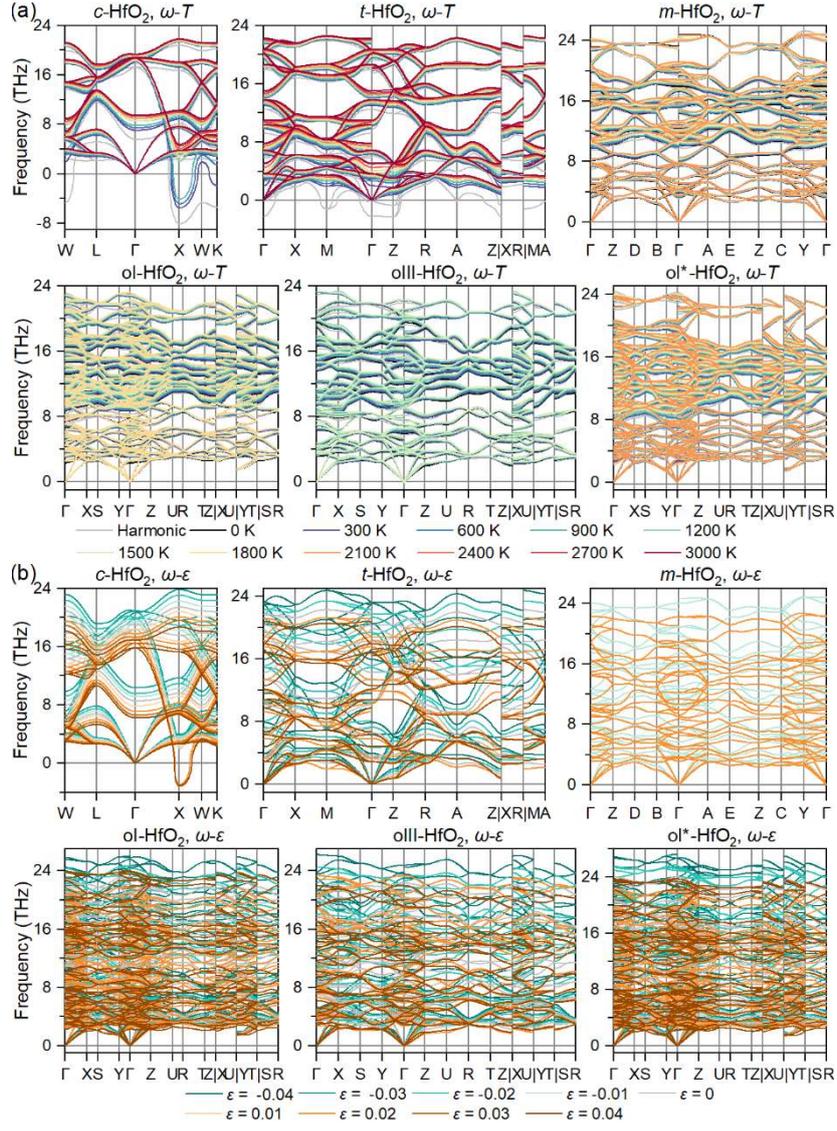

**Figure 2.** (a) SCP spectra of a given *c*-, *t*-, *m*-, oI- and oIII-HfO$_2$ lattice over temperature. (b) SCP spectra of *c*-, *t*-, *m*-, oI- and oIII-HfO$_2$ at 1200 K over equiaxial strain with respect to DFT-optimized lattice.

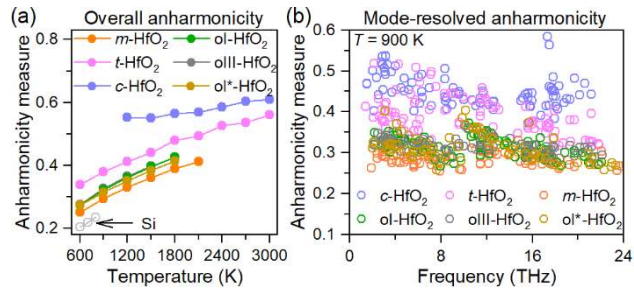

**Figure 3.** (a) Overall anharmonicity measure of *c*-, *t*-, *m*-, oI- and oIII-HfO$_2$ as a function of temperature. Results for Si are extracted from ref. 40. (b) Mode-resolved anharmonicity measure of *c*-, *t*-, *m*-, oI- and oIII-HfO$_2$ at 900 K.

To validate the above SCP and thermodynamic calculations, we compared the simulated temperature and pressure dependency of the thermodynamic quantities. The pressure-dependence of the normalized unit-cell volume is shown in Figure 4a, which consists with the experimental results of oI-HfO$_2$[7] and m-HfO$_2$.[41,42] In addition, the volumetric thermal expansion of m-, t- and c-HfO$_2$ (Figure 4b) qualitatively coincides with experiments.[43,44] Based on the validated simulation of thermodynamics, the Gibbs free energy of each phase is interpolated via third-order polynomial. Typically, we showcase the resulting Gibbs free energy profile at 1200 K in Figure 4c, where m-HfO$_2$ (oI-HfO$_2$) is ground state below (over) 2.3 GPa. The minimization of Gibbs free energy over the simulated phases yields the p-T phase diagram of HfO$_2$ shown in Figure 4d, where m-HfO$_2$ and antipolar oI-HfO$_2$ are respectively identified as the low-temperature ground state at zero and low pressure, and t-HfO$_2$ is identified as the ground state at higher temperature, quantitatively consistent with the experiments.[7]

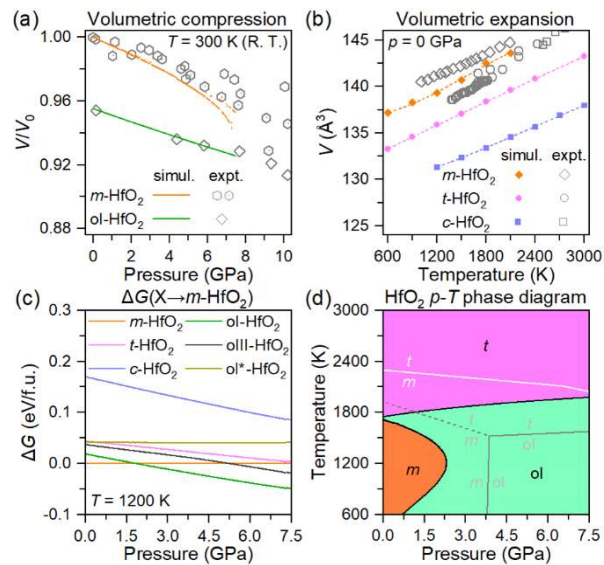

**Figure 4.** (a) Room-temperature unit-cell volume of m- and oI-HfO$_2$ (normalized by the zero-pressure unit-cell volume of m-HfO$_2$) as a function of pressure. Experimental results are extracted from refs. 7, 41, 42. (b) Zero pressure unit-cell volume of m- and oI-HfO$_2$ as a function of temperature. Dashed lines are polynomial fit of raw data to guide the eye. Experimental

results are extracted from refs. 43, 44. (c) Relative Gibbs free energy with respect to $m$-HfO$_2$ at 1200 K. (d) Simulated $p$-$T$ phase diagram of HfO$_2$, where solid (dashed) black lines represent stable (metastable) phase border, solid (dashed) grey lines represent experimental confirmed (assumed) phase borders extracted from ref. 7 and solid white line represent simulated harmonic-level phase border extracted from ref. 45.

**Low-temperature metastable ferroelectric region with potentially temperature-dependent parent phase.** Previous simulation based on harmonic model and Helmholtz free energy[45] predicts that the metastable region defined by $\Delta G \leq 0.2 k_B T$ is found above the line ($T$ = 1500 K, $p$ = 0)-($T$ = 500 K, $p$ = 10 GPa), trivially covering the three-phase point of $t$-, $m$- and oI-HfO$_2$ where all these phases possess similar Gibbs free energy. However, one can see that the Gibbs free energy difference $\Delta G$ between the metastable ferroelectric oIII-HfO$_2$ and the ground state, as a measure of the metastability, is generally lower than $0.1 k_B T$ throughout the $m$- and oI-HfO$_2$ phase regions, as represented by the greenish regions in Figure 5a, qualitatively lower than the previous prediction in this region (generally $\Delta G > 0.2 k_B T$).[45] In fact, despite the large static energy difference $\Delta E$ among phases, the Gibbs free energy difference $\Delta G$ is always much smaller, as shown in Figure 5b. The metastable energy profile of oIII-HfO$_2$ leads to the formation of $m$- and oI-HfO$_2$ phases in the ferroelectric devices.[8] In the meantime, the unexpectedly low Gibbs free energy difference between metastable oIII-HfO$_2$ and the ground state phases at low temperatures helps explain the dominating ferroelectric flipping in devices despite the thermodynamic instability of oIII-HfO$_2$, holding out the prospect of achieving ground-state oIII-HfO$_2$ *via* doping.

In addition to the overhauled metastable region of oIII-HfO$_2$, the SCP-level thermodynamics could also provide insight into the everlasting controversy in the true parent phase of ferroelectric oIII-$HfO_2$. Take oI*-$HfO_2$ as example, despite its lower static energy as compared to the intuitive parent phase $t$-$HfO_2$, its higher vibrational free energy gradually cancels such merit over increasing temperature. As shown in Figure 5c, typically, oI*-$HfO_2$ is thermodynamically more stable than $t$-$HfO_2$ at low temperature and pressure (orange region), which, by extrapolation, points to the likely stability of oI*-$HfO_2$ over $t$-$HfO_2$ at ambient conditions. Accordingly, the "best" parent phase should naturally be expected as a temperature- and pressure-dependent entity rather than a universal single phase. Given the polymorphous nature of the statistics over a flat potential energy surface,[46] where $t$-$HfO_2$ is essentially the ensemble average over the rattled configurations hosted by several neighboring local minima in the configuration space, $t$-$HfO_2$ could be the only rational parent phase at sufficiently high temperature despite its high zero-temperature static energy over all rivals.

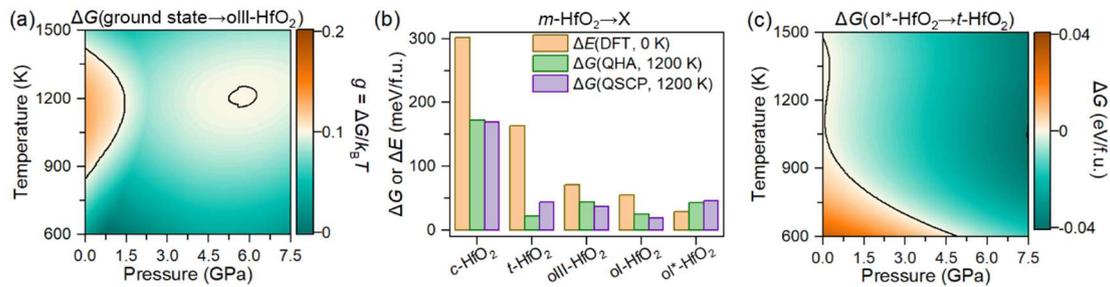

**Figure 5.** (a) Gibbs free energy difference between ferroelectric oIII-$HfO_2$ and the ground state renormalized by average atomic kinetic energy. (b) Phase-resolved Gibbs free energy at 1200 K and zero pressure referenced by that of the ground-state $m$-$HfO_2$. (c) Gibbs free energy difference between potential parent phases oI*- and $t$-$HfO_2$.

In summary, based on first-principles thermodynamic calculations facilitated by equivariant MLFF in the MACE architecture, we qualitatively reproduce the $p$-$T$ phase diagram of $HfO_2$ from the phase-resolved Gibbs free energy profile and analyzed the underlying phase-equilibrium thermodynamics. Anharmonicity proves profound in all simulated $HfO_2$ phases such that it plays a crucial part in thermodynamics. The simulated low-lying metastable region of oIII-$HfO_2$ ($\Delta G < 0.1k_BT$ as compared to ground state) helps explain the emergence of *m*- and oI-$HfO_2$ in metastable oIII-$HfO_2$ upon its relatively robust ferroelectric cycles, indicating the thermodynamic stabilization of ferroelectric oIII-$HfO_2$ over these phases as key to durable ferroelectric devices. The Gibbs free energy inversion between the obvious high-static-energy ferroelectric parent phase *t*-$HfO_2$ and a delicate low-static-energy rival oI*-$HfO_2$ highlights the essence of thermodynamics in ferroelectric switching mechanism. This study showcases the technical ground for the rational design of $HfO_2$-based ferroelectric devices.

**Acknowledgements**

This work was supported by National Natural Science Foundation of China (Grant No. 52003150, 52130204), the Program for Young Eastern Scholar at Shanghai Institutions of Higher Education (Grant No. QD2019006), the Science and Technology Commission of Shanghai Municipality (No. 25511103400 and 24CL2901702), and Shanghai Technical Service Computing Center of Science and Engineering, Shanghai University.

# Supporting Information

**S1. Additional computational methods for DFT**

All first-principles calculations were performed by using the Vienna *ab initio* simulations package (VASP)[1,2] based on DFT, where the projector augmented wave method[3] was used for describing the interaction between valence electrons and ion cores, and the plane waves with an energy cutoff of 600 eV were used to expand the electron wave functions. The exchange-correlation interaction was treated with the SCAN functional[4] for all calculations except those for the Born effective charges, which were treated with the PBEsol functional[5]. The integration over the first Brillouin zone was performed on a generalized regular **k**-point grid[6,7] with the grid density controlled by KSPACING set to 0.025. The convergence thresholds for the total energy during the self-consistent field iteration and the interatomic forces during the geometry optimization were respectively set to $10^{-5}$ eV and $10^{-3}$ eV Å$^{-1}$. First principles calculations on eight solid phases (*c*, *t*, *m*, *r*, oI, oII, oIII, and oIV) of $HfO_2$ were performed, as shown in Figure S1 and summarized in Table S1, to provide data for developing MLFF. Thermodynamic calculations are performed on the four ground-state phases in the *p-T* phase-diagram (*c*, *t*, *m* and oI) in addition to the oIII phase.

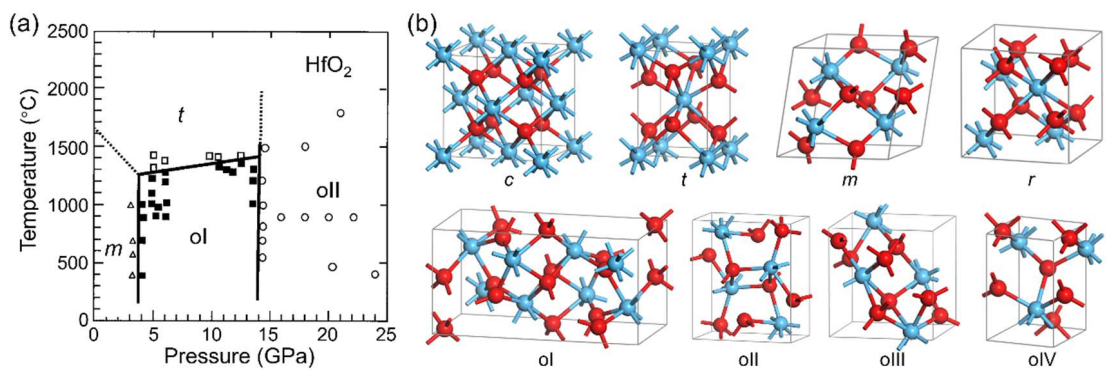

**Figure S1.** (a) Experimental *p-T* phase diagram of $HfO_2$ extracted from ref. 8. (b) Unit cells of the solid phases of $HfO_2$ concerned in this work.

**Table S1** Symmetry and polarity of the solid phases of $HfO_2$ concerned in this work.

| System | Crystal system | Space group | Polarity | DFT calc. | SCP calc. |
|---|---|---|---|---|---|
| *c*-$HfO_2$ | Cubic | $Fm\bar{3}m$ (no. 225) | Paraelectric | √ | √ |
| *t*-$HfO_2$ | Tetragonal | $P4_2/mnc$ (no. 137) | Paraelectric | √ | √ |
| *m*-$HfO_2$ | Monoclinic | $P2_1/c$ (no. 14) | Paraelectric | √ | √ |
| *r*-$HfO_2$ | Rhombohedral | $R3m$ (no. 160) | Ferroelectric | √ | × |
| oI-$HfO_2$ | Orthorhombic | $Pbca$ (no. 61) | Antipolar | √ | √ |
| oII-$HfO_2$ | Orthorhombic | $Pnma$ (no. 62) | Paraelectric | √ | × |
| oIII-$HfO_2$ | Orthorhombic | $Pca2_1$ (no. 29) | Ferroelectric | √ | √ |
| oIV-$HfO_2$ | Orthorhombic | $Pmn2_1$ (no. 31) | Ferroelectric | √ | × |
| oI*-$HfO_2$ | Orthorhombic | $Pbca$ (no. 61) | Paraelectric | × | √ |

**S2. Additional computational methods for SCP**

The SCP calculations were performed by stochastic temperature-dependent effective potential (sTDEP) using self_consistent_harmonic_model function in the hiPhive package[9] with necessary modifications to implement the non-analytical correction (NAC).[10,11] In brief, the columbic forces are subtracted from the total interatomic forces prior to the least-square fitting of effective harmonic model, and the columbic force constants are added back to the fitted effective harmonic model prior to the generation of phonon-rattled configuration. As compared to the original implementation, the modified version scph_mlff, as provided in the following box, additionally takes in the columbic force constants fc2_LR, which can be generated from Born effective charges using the Phonopy package[12] as showcased in the HiPhive document (Advanced topics –> Non-analytic term correction,

https://hiphive.materialsmodeling.org/advanced_topics/long_range_forces.html#Non-analytic-term-correction). All necessary modifications for processing forces, force constants and reference energy $U_0$ in accordance with NAC are marked red. Here, reference energy $U_0$ is the first-order perturbation of the effective harmonic potential with respect to the true potential energy surface[13] represented by MLFF, and the total Helmholtz free energy $F$ at given temperature and lattice parameters is calculated as the sum of vibrational Helmholtz free energy $F_{vib}$ from SCP plus $U_0$.

```python
import numpy as np
import copy
from hiphive import ForceConstantPotential, ClusterSpace, StructureContainer,\
                    enforce_rotational_sum_rules
from hiphive.force_constants import ForceConstants
from hiphive.force_constant_model import ForceConstantModel
from hiphive.calculators import ForceConstantCalculator
from hiphive.structure_generation import generate_rattled_structures, \
                                         generate_phonon_rattled_structures
from trainstation import Optimizer
from hiphive.utilities import prepare_structures
from hiphive.input_output.logging_tools import set_config

def compute_ave_ener(structures, calc):
    E = []
    for structure in structures:
        structure.set_calculator(calc)
    return np.mean([structure.get_potential_energy() for structure in structures])

def scph_mlff(atoms_ideal, calc, cs, T, alpha, n_iterations, n_structures, \
              parameters_start=None, fc2_LR=None, fit_kwargs={}):
    if not 0 < alpha <= 1:
        raise ValueError('alpha must be between 0.0 and 1.0')
    if max(cs.cutoffs.orders) != 2:
        raise ValueError('ClusterSpace must be second order')
    set_config(level=40)

    # initialize things
    scph_log = []
```

```python
        sc = StructureContainer(cs)
        fcm = ForceConstantModel(atoms_ideal, cs)
        if not (fc2_LR is None):
            fc2_LR_ase = fc2_LR.transpose(0, 2, 1, 3). \
                                  reshape(3 * len(atoms_ideal), 3 * len(atoms_ideal))
            fc2_LR_phon = fc2_LR
            calc_nac = ForceConstantCalculator(ForceConstants.from_arrays(atoms_ideal, \
                                         fc2_LR_phon), max_disp=20.0)

        # generate initial model
        if parameters_start is None:
            print('Creating initial model', flush=True)
            rattled_structures = generate_rattled_structures(atoms_ideal, n_structures, 0.03)
            rattled_structures = prepare_structures(rattled_structures, atoms_ideal, calc, False)
            for structure in rattled_structures:
                sc.add_structure(structure)
            M, F = sc.get_fit_data()
            if not (fc2_LR is None):
                displacements = np.array([structure.displacements for structure in sc])
                F -= np.einsum('ijab,njb->nia', -fc2_LR_phon, displacements).flatten()
            opt = Optimizer((M, F), train_size=1.0, **fit_kwargs)
            opt.train()
            parameters_start = enforce_rotational_sum_rules(cs, opt.parameters, \
                                                             ['Huang','Born-Huang'])
            sc.delete_all_structures()

        # run poor mans self consistent
        parameters_old = parameters_start.copy()
        parameters_traj = [parameters_old]

        for i in range(n_iterations):
            # generate structures with old model
            print('Iteration {}'.format(i), flush=True)
            fcm.parameters = parameters_old
            fc2 = fcm.get_force_constants().get_fc_array(order=2, format='ase')
            if not (fc2_LR is None):
                fc2 += fc2_LR_ase
            phonon_rattled_structures =\
                    generate_phonon_rattled_structures(atoms_ideal, fc2, n_structures,\
                                                         T, QM_statistics=True)
            calc_old = ForceConstantCalculator(ForceConstantPotential(cs, \
                        parameters_old).get_force_constants(atoms_ideal), max_disp=20.0)
            E_old = compute_ave_ener(phonon_rattled_structures, calc_old)
            E_true = compute_ave_ener(phonon_rattled_structures, calc)
```

```python
            if not (fc2_LR is None):
                E_nac = compute_ave_ener(phonon_rattled_structures, calc_nac)
                U0 = E_true - E_old - E_nac
            else:
                U0 = E_true - E_old
            for structure in phonon_rattled_structures:
                structure.set_calculator(None)
            phonon_rattled_structures = prepare_structures(phonon_rattled_structures, \
                                                      atoms_ideal, calc, False)

            # fit new model
            for structure in phonon_rattled_structures:
                sc.add_structure(structure)
            M, F = sc.get_fit_data()
            if not (fc2_LR is None):
                displacements = np.array([structure.displacements for structure in sc])
                F -= np.einsum('ijab,njb->nia', -fc2_LR_phon, displacements).flatten()
            opt = Optimizer((M, F), train_size=1.0, **fit_kwargs)
            opt.train()
            sc.delete_all_structures()

            # update parameters
            parameters_new = alpha * np.array(opt.parameters) + \
                             np.array((1-alpha)) * parameters_old

            # print iteration summary
            disps = [atoms.get_array('displacements') for atoms in phonon_rattled_structures]
            disp_ave = np.mean(np.abs(disps))
            disp_max = np.max(np.abs(disps))
            x_new_norm = np.linalg.norm(parameters_new)
            delta_x_norm = np.linalg.norm(parameters_old-parameters_new)
            print('     |x_new| = {:.5f}, |delta x| = {:.8f}, disp_ave = {:.5f}, disp_max = {:.5f}, '
                  'rmse = {:.5f}'.format(x_new_norm, delta_x_norm, disp_ave, disp_max, \
                                      opt.rmse_train), flush=True)
            if not (fc2_LR is None):
                scph_log.append([U0, E_true, E_old, E_nac, x_new_norm, delta_x_norm, \
                              disp_ave, disp_max, opt.rmse_train])
            else:
                scph_log.append([U0, E_true, E_old, x_new_norm, delta_x_norm, \
                              disp_ave, disp_max, opt.rmse_train])
            parameters_traj.append(parameters_new)
            parameters_old = parameters_new

        # evaluate last model
```

```
    fcm.parameters = parameters_old
    fc2 = fcm.get_force_constants().get_fc_array(order=2, format='ase')
    if not (fc2_LR is None):
        fc2 += fc2_LR_ase
    phonon_rattled_structures = generate_phonon_rattled_structures(atoms_ideal, fc2, \
                                                    n_structures, T, QM_statistics=True)

    calc_old = ForceConstantCalculator(ForceConstantPotential(cs, \
                    parameters_old).get_force_constants(atoms_ideal), max_disp=20.0)
    E_old = compute_ave_ener(phonon_rattled_structures, calc_old)
    E_true = compute_ave_ener(phonon_rattled_structures, calc)
    if not (fc2_LR is None):
        E_nac = compute_ave_ener(phonon_rattled_structures, calc_nac)
        U0 = E_true - E_old - E_nac
        scph_log.append([U0, E_true, E_old, E_nac])
    else:
        U0 = E_true - E_old
        scph_log.append([U0, E_true, E_old])

    return parameters_traj, scph_log
```

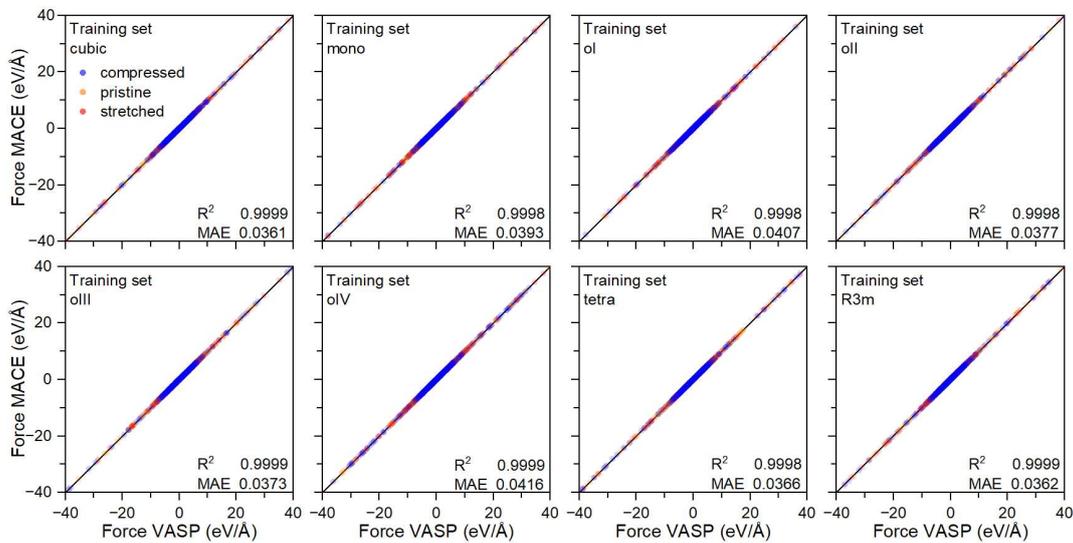

**Figure S2.** Parity plot for the structure- and strain-resolved interatomic forces on the training set calculated by VASP and the MACE MLFF.

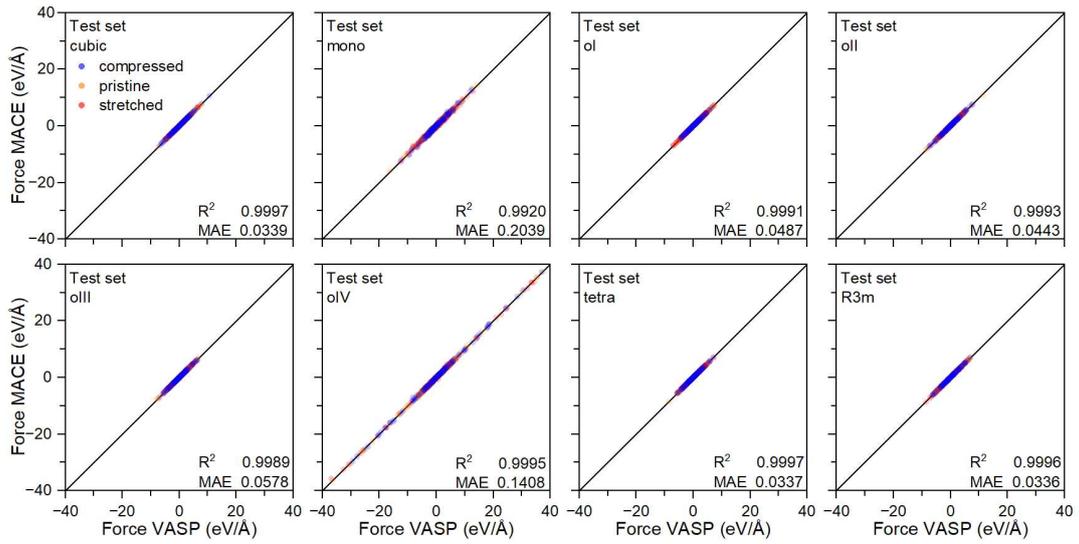

**Figure S3.** Parity plot for the structure- and strain-resolved interatomic forces on the test set calculated by VASP and the MACE MLFF.

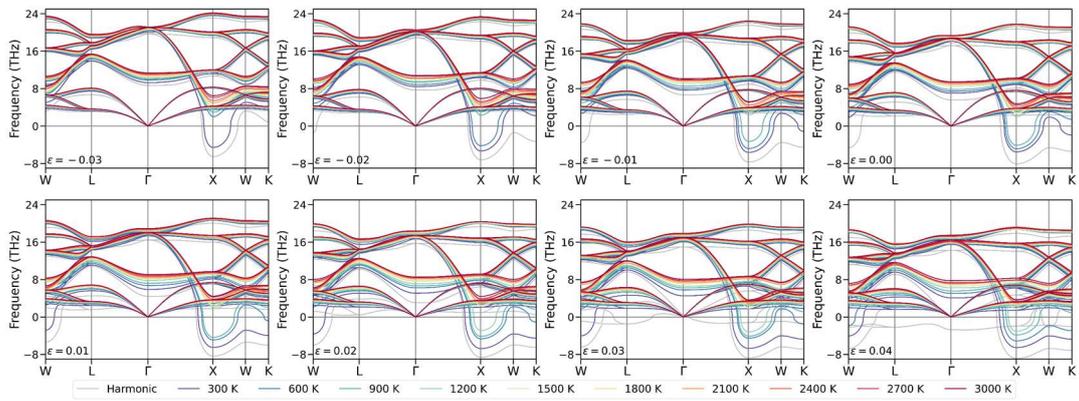

**Figure S4.** SCP spectra of $c$-HfO$_2$ under equiaxial strain $\varepsilon$.

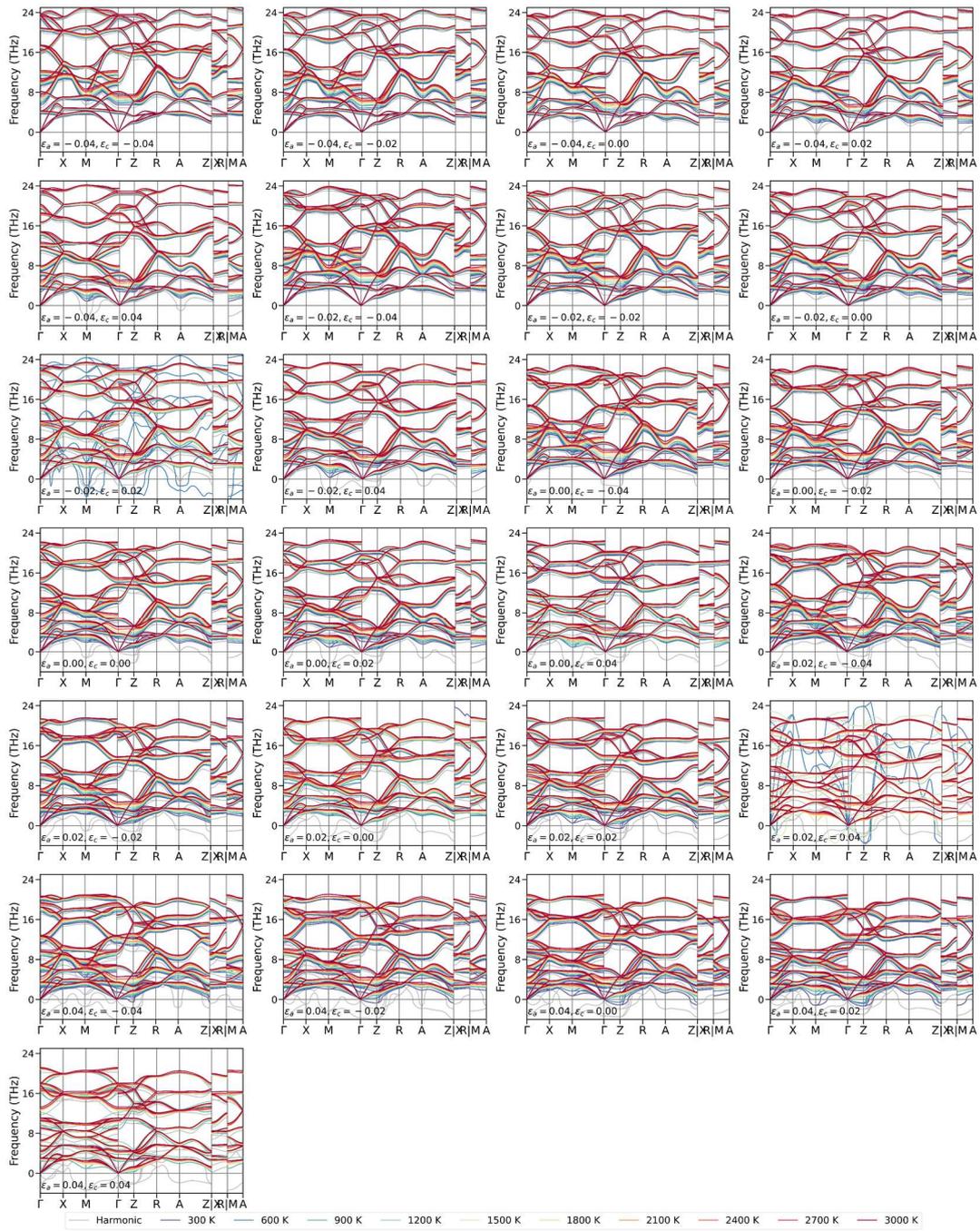

**Figure S5.** SCP spectra of *t*-HfO$_2$ under biaxial strain $\varepsilon_a$ and uniaxial strain $\varepsilon_c$. Missing or gibberish curves represent diverged SCP calculations.

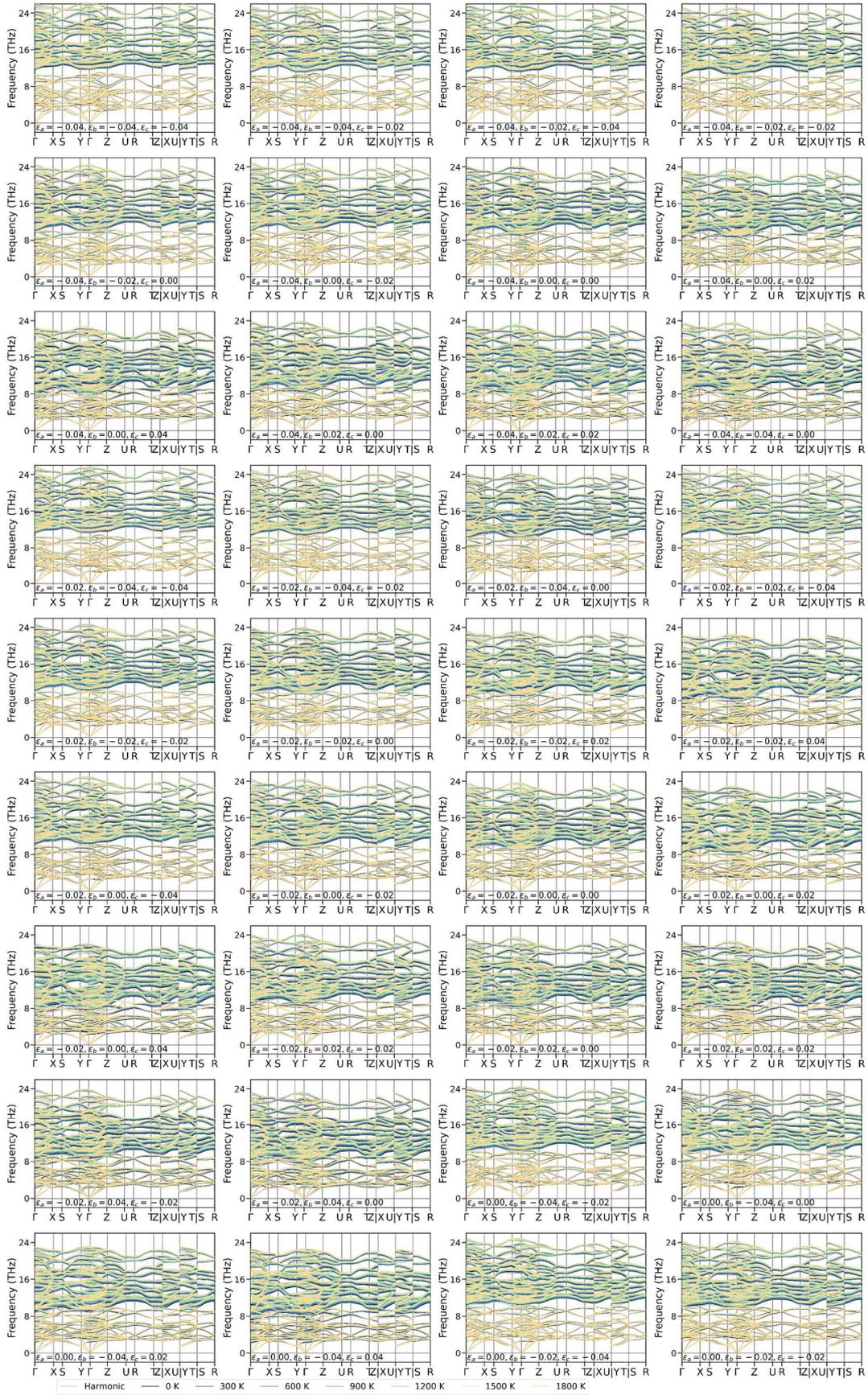

**Figure S6.** SCP spectra of oI-HfO$_2$ under uniaxial strains $\varepsilon_a$, $\varepsilon_b$ and $\varepsilon_c$.

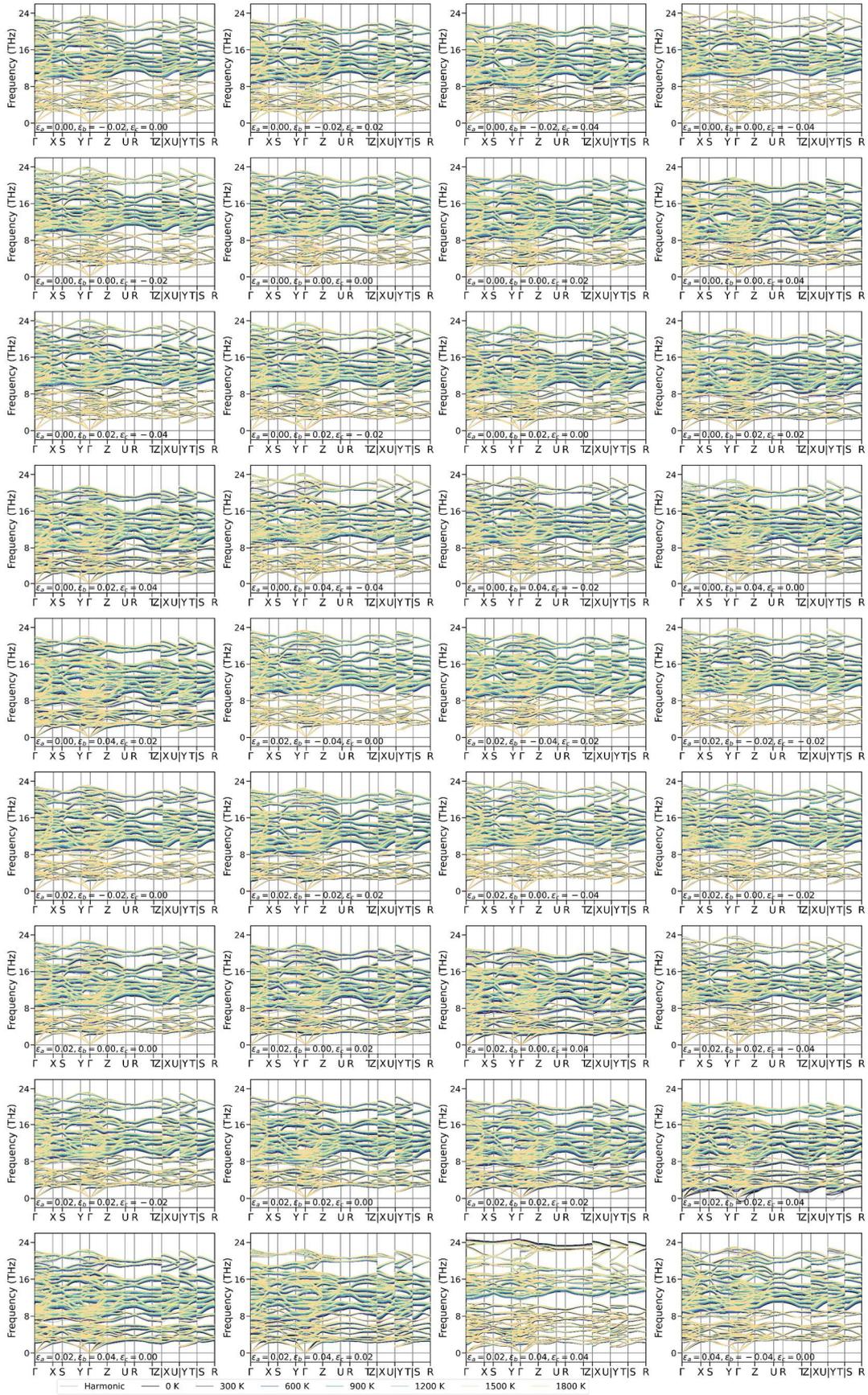

**Figure S6.** (continue) SCP spectra of oI-HfO$_2$ under uniaxial strains $\varepsilon_a$, $\varepsilon_b$ and $\varepsilon_c$. 6

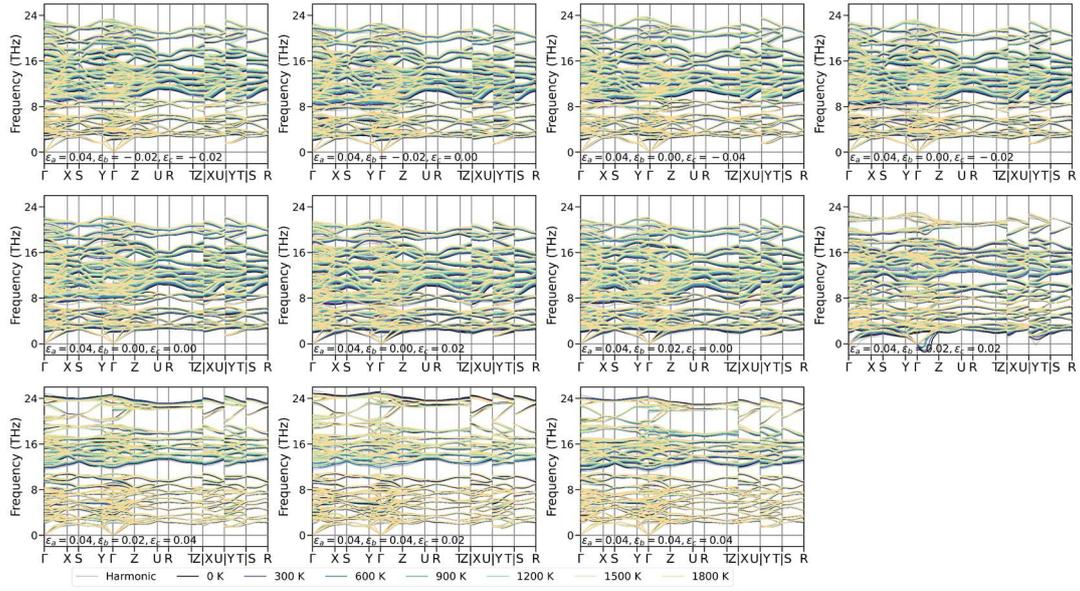

**Figure S6.** (continue) SCP spectra of oI-HfO$_2$ under uniaxial strains $\varepsilon_a$, $\varepsilon_b$ and $\varepsilon_c$.

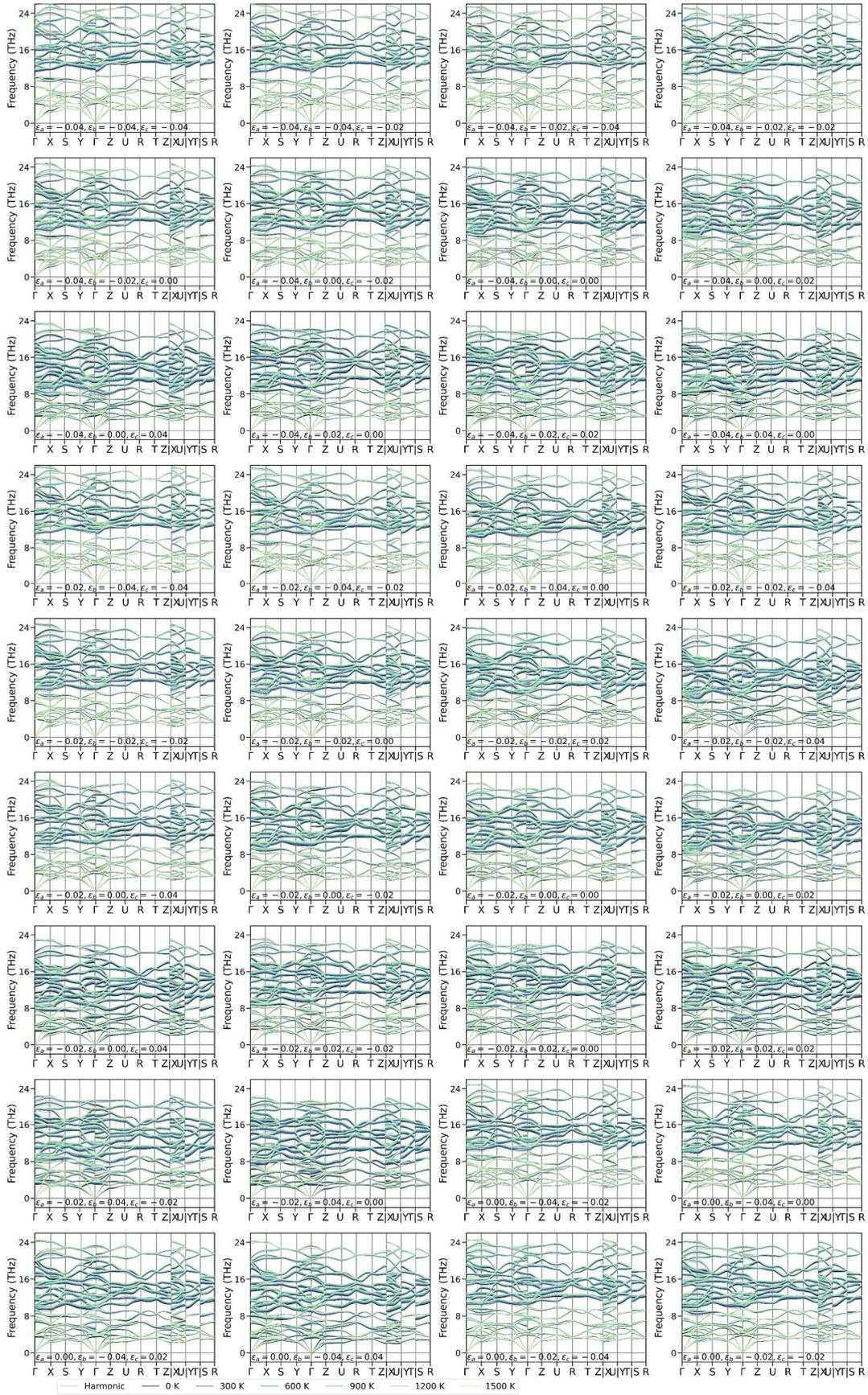

**Figure S7.** SCP spectra of oIII-HfO$_2$ under uniaxial strains $\varepsilon_a$, $\varepsilon_b$ and $\varepsilon_c$.

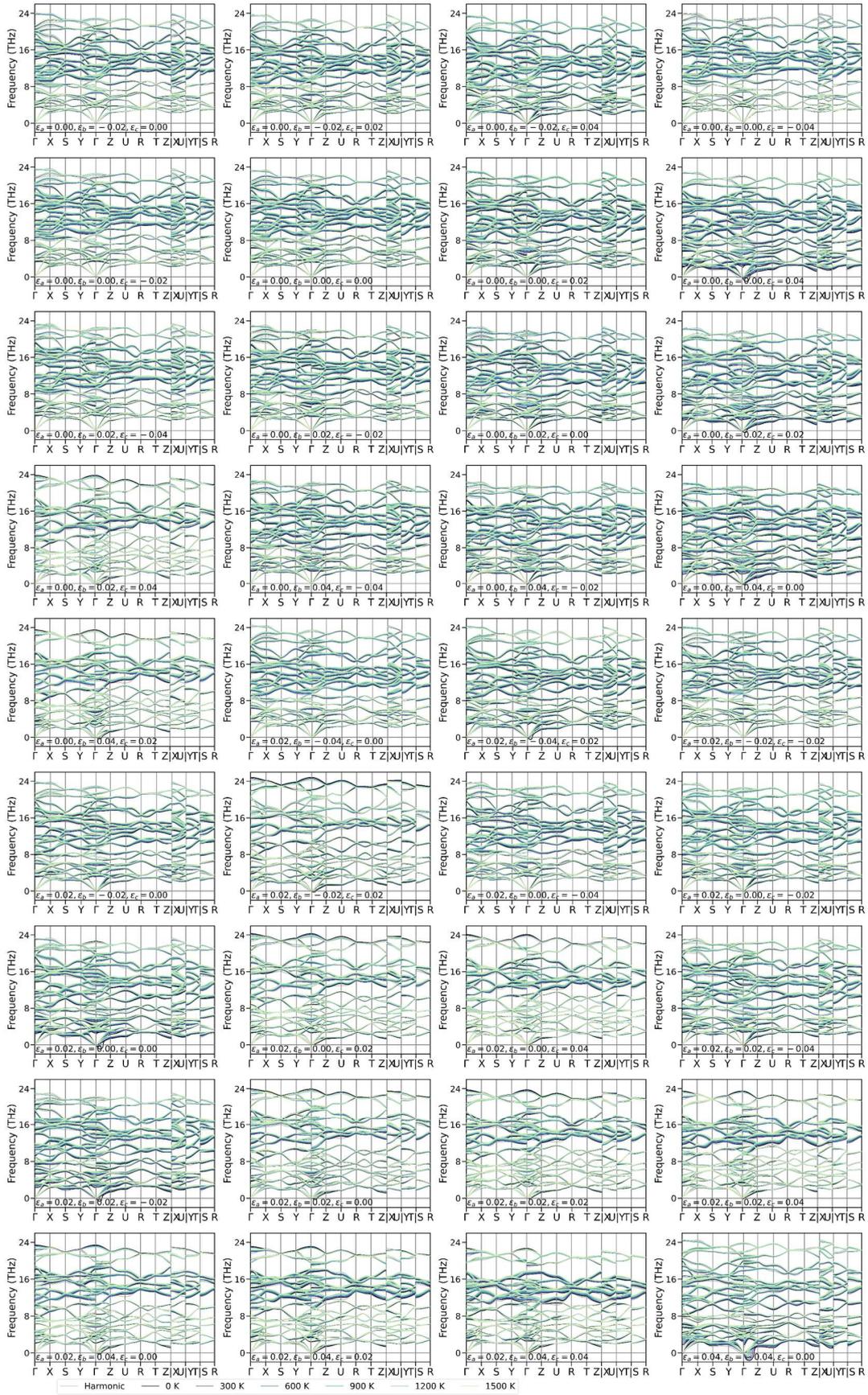

**Figure S7.** (continue) SCP spectra of oIII-HfO$_2$ under uniaxial strains $\varepsilon_a$, $\varepsilon_b$ and $\varepsilon_c$.

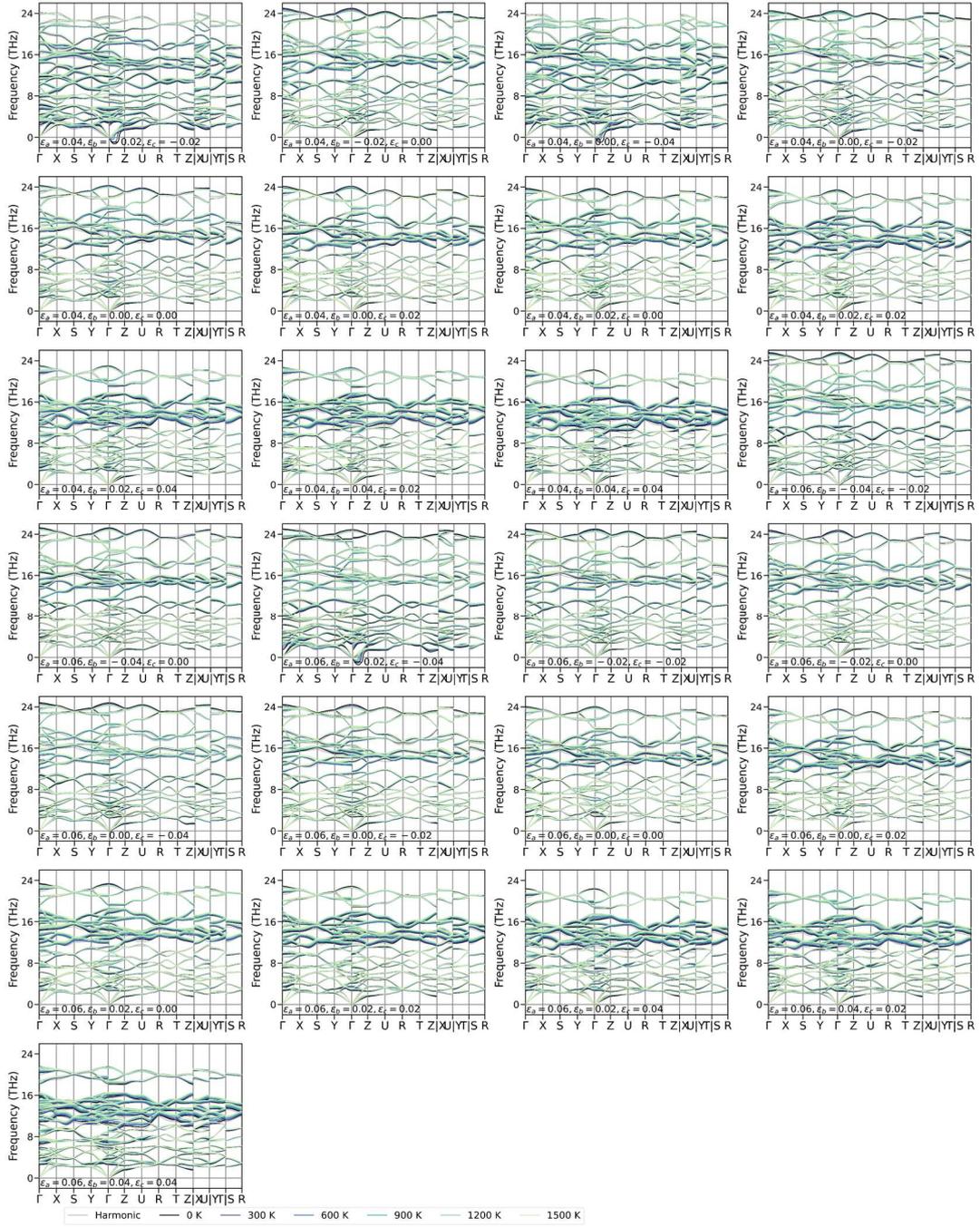

**Figure S7.** (continue) SCP spectra of oIII-HfO$_2$ under uniaxial strains $\varepsilon_a$, $\varepsilon_b$ and $\varepsilon_c$.

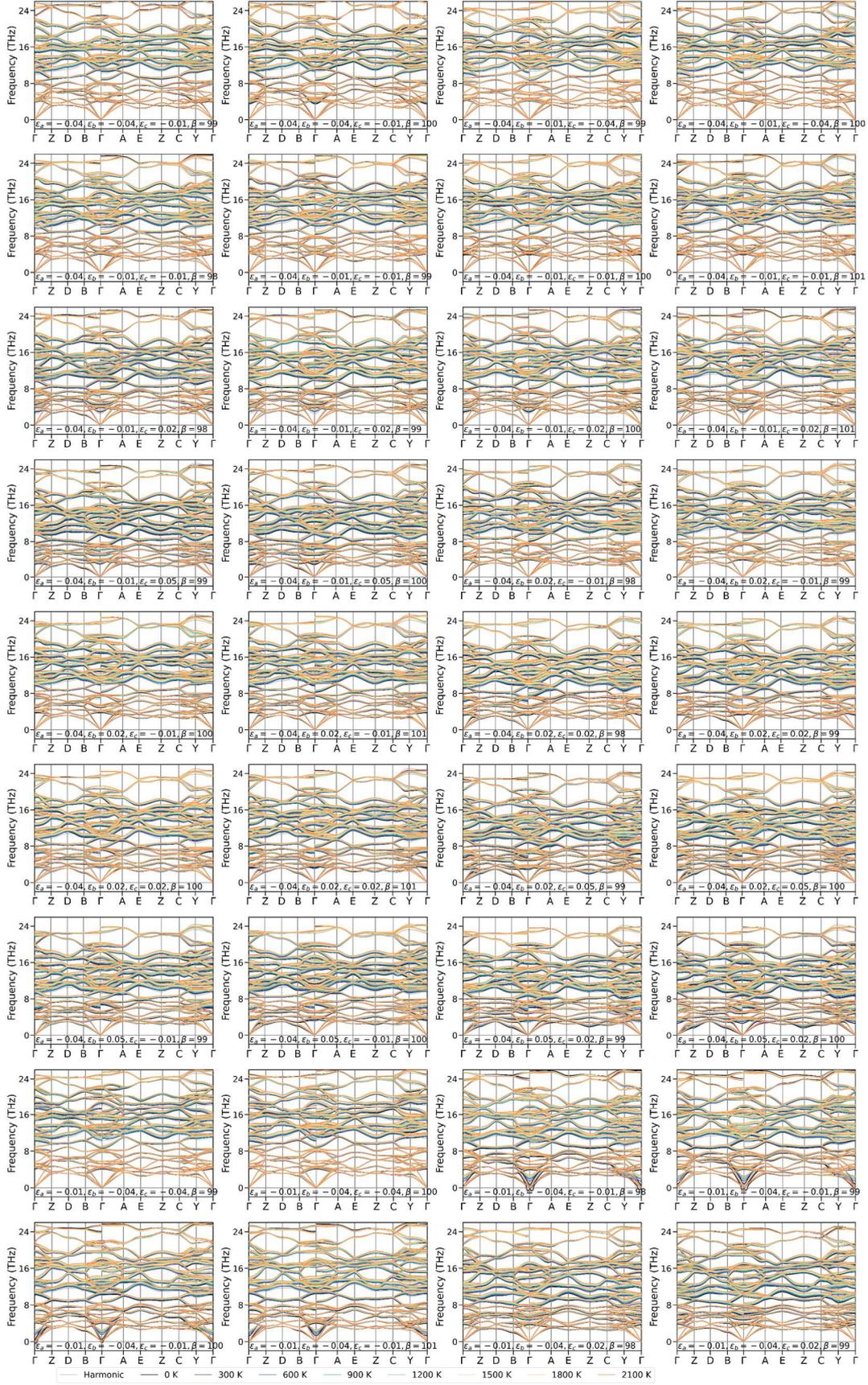

**Figure S8.** SCP spectra of *m*-HfO$_2$ under uniaxial strains $\varepsilon_a$, $\varepsilon_b$ and $\varepsilon_c$ and different lattice

parameter $\beta$.

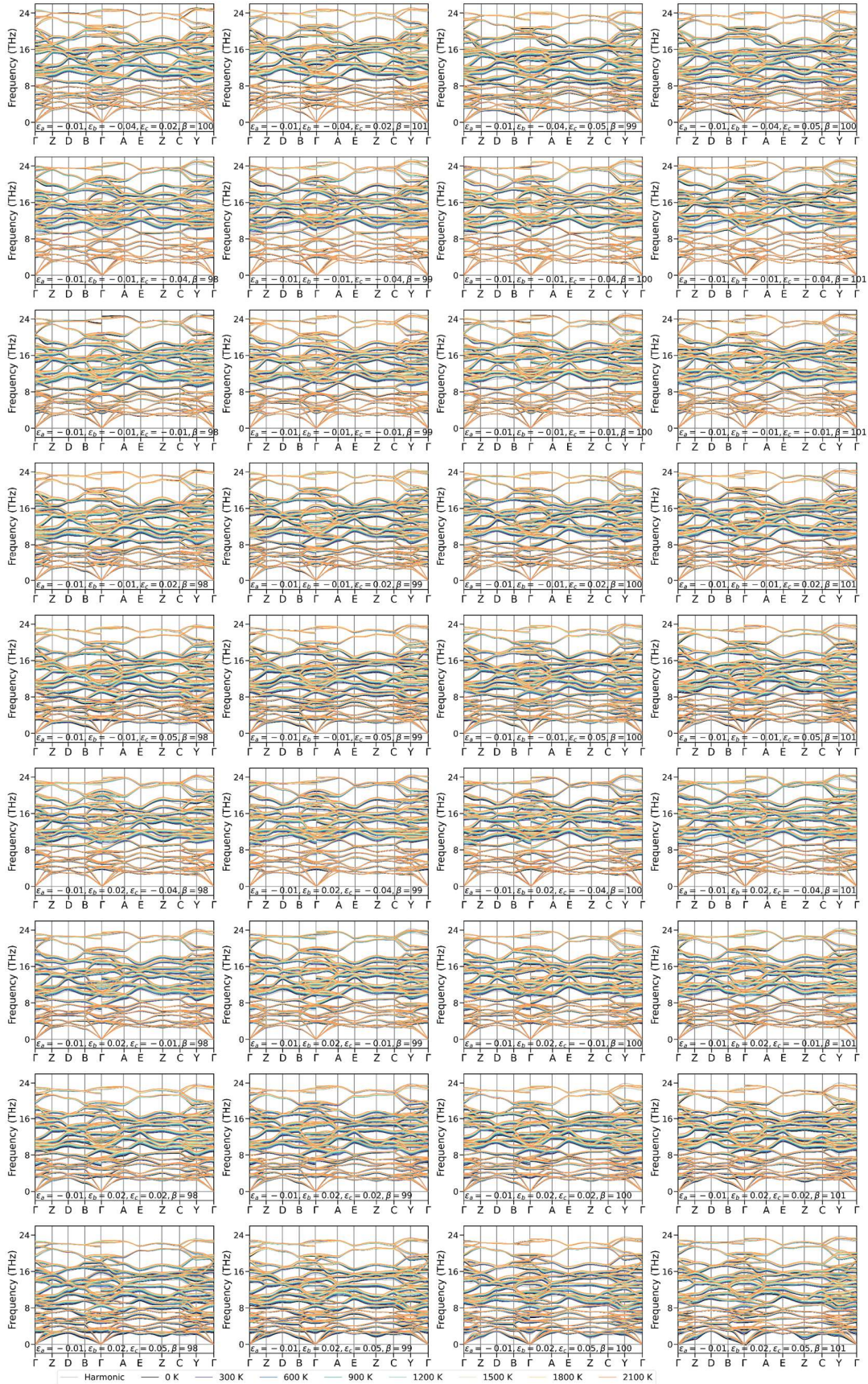

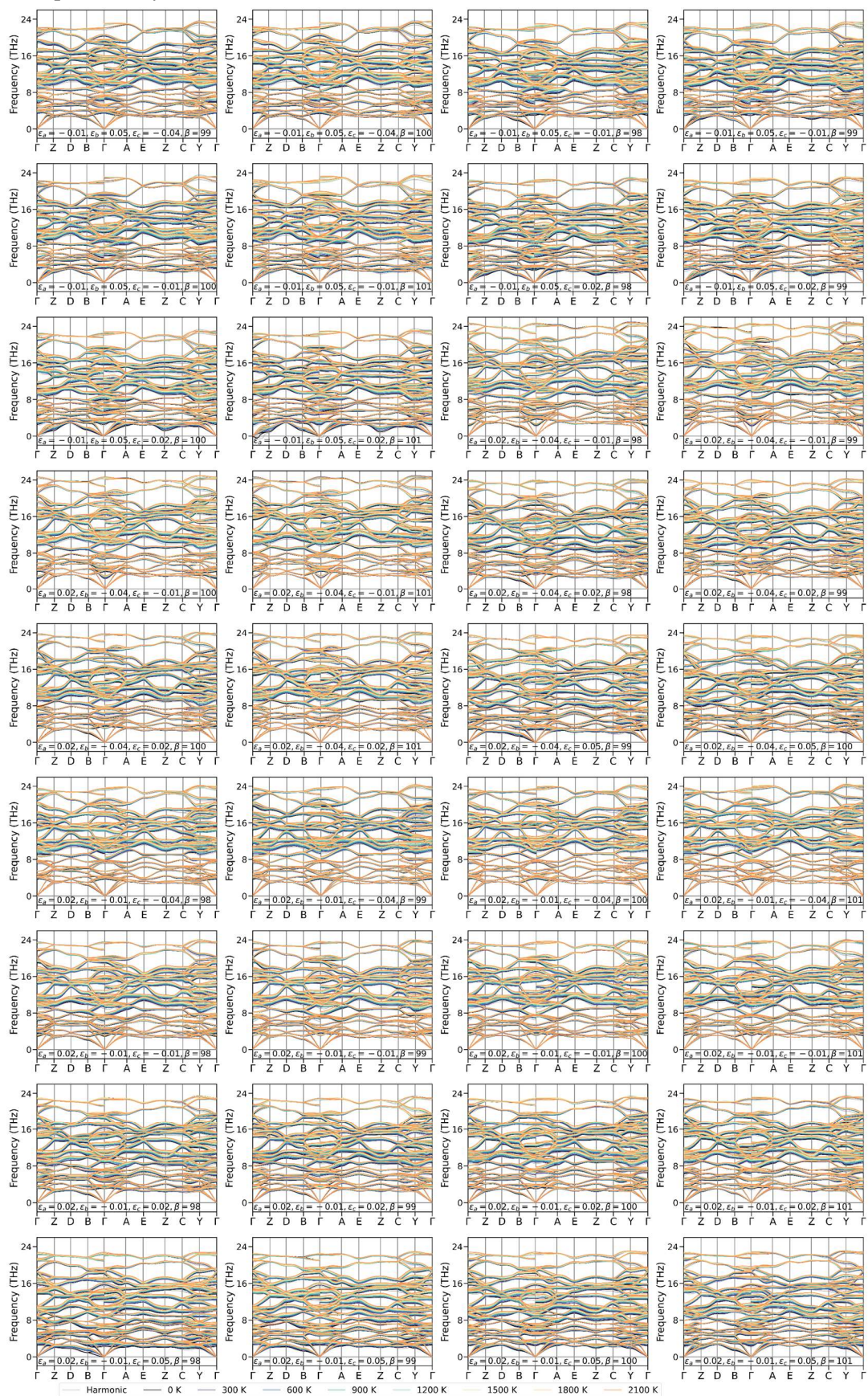

**Figure S8.** (continue) SCP spectra of *m*-HfO$_2$ under uniaxial strains $\varepsilon_a$, $\varepsilon_b$ and $\varepsilon_c$ and different lattice parameter $\beta$.

**Figure S8.** (continue) SCP spectra of *m*-HfO$_2$ under uniaxial strains $\varepsilon_a$, $\varepsilon_b$ and $\varepsilon_c$ and different lattice parameter $\beta$.

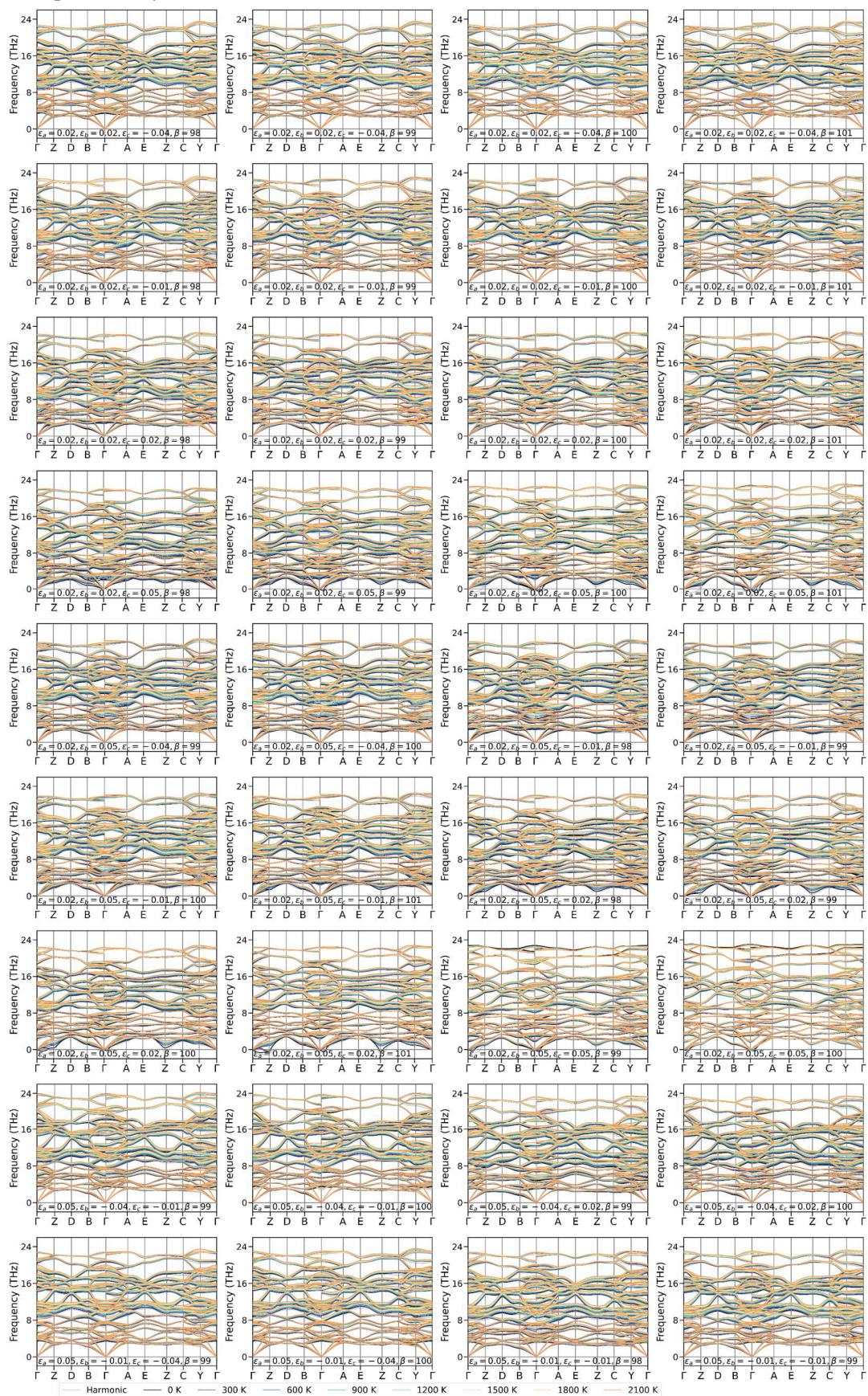

**Figure S8.** (continue) SCP spectra of *m*-HfO$_2$ under uniaxial strains $\varepsilon_a$, $\varepsilon_b$ and $\varepsilon_c$ and different lattice parameter $\beta$.

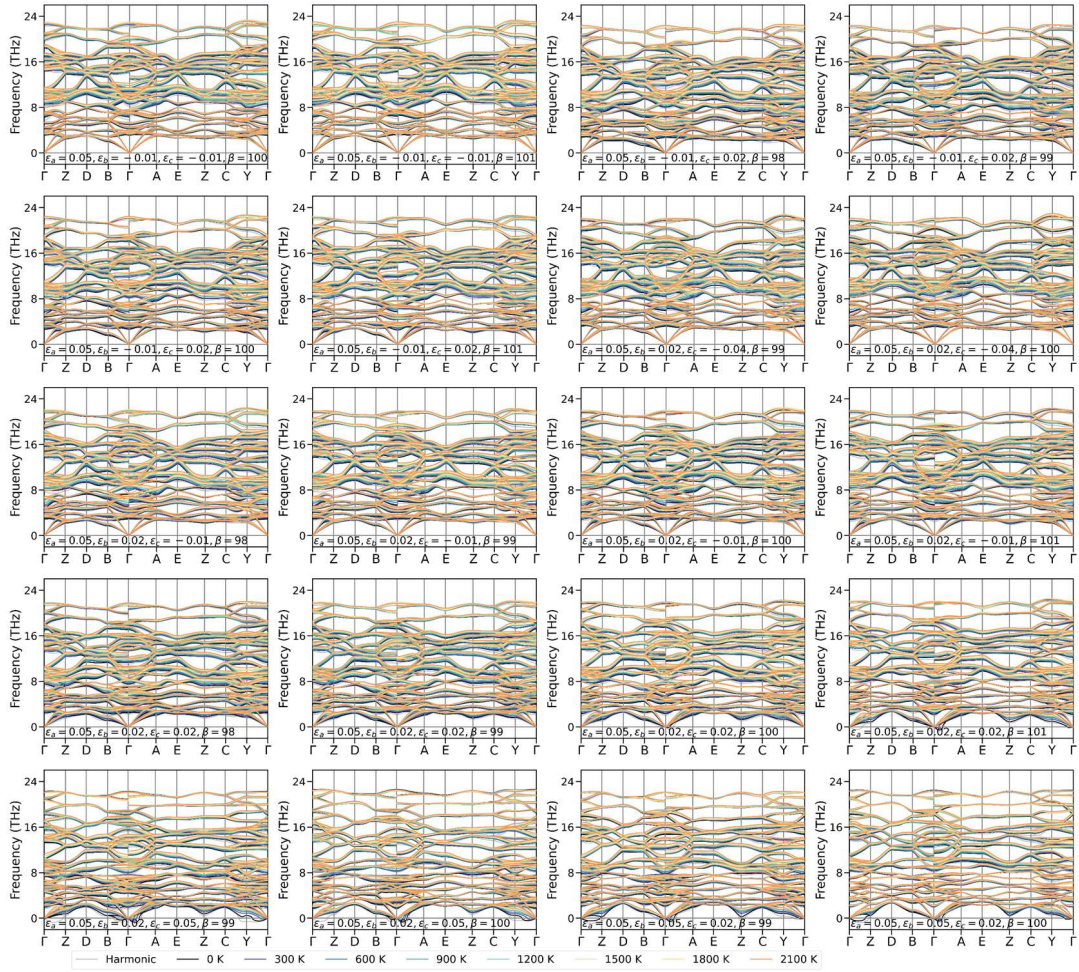

**Figure S8.** (continue) SCP spectra of *m*-HfO$_2$ under uniaxial strains $\varepsilon_a$, $\varepsilon_b$ and $\varepsilon_c$ and different lattice parameter $\beta$.

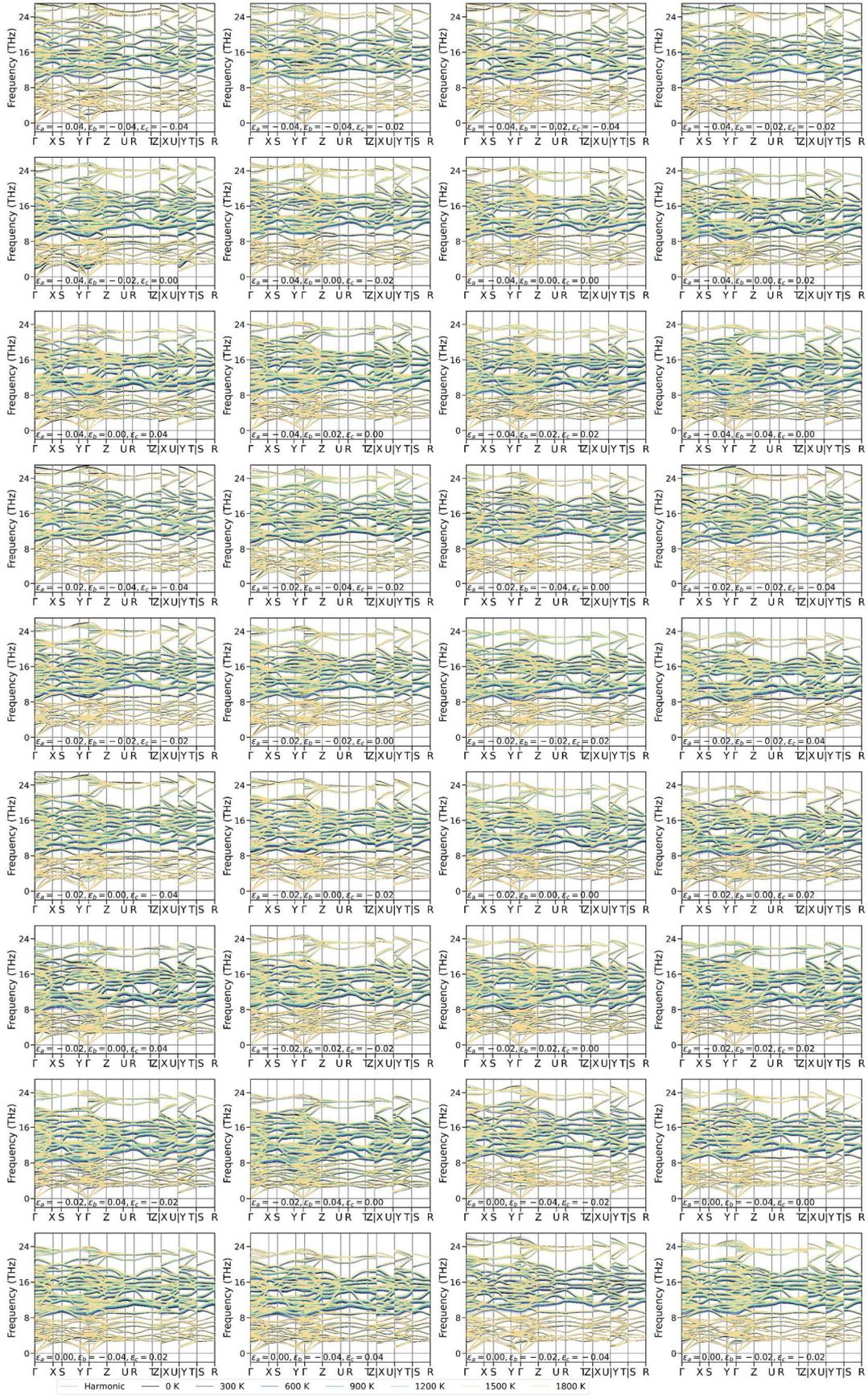

**Figure S9.** SCP spectra of oI*-HfO$_2$ under uniaxial strains $\varepsilon_a$, $\varepsilon_b$ and $\varepsilon_c$.

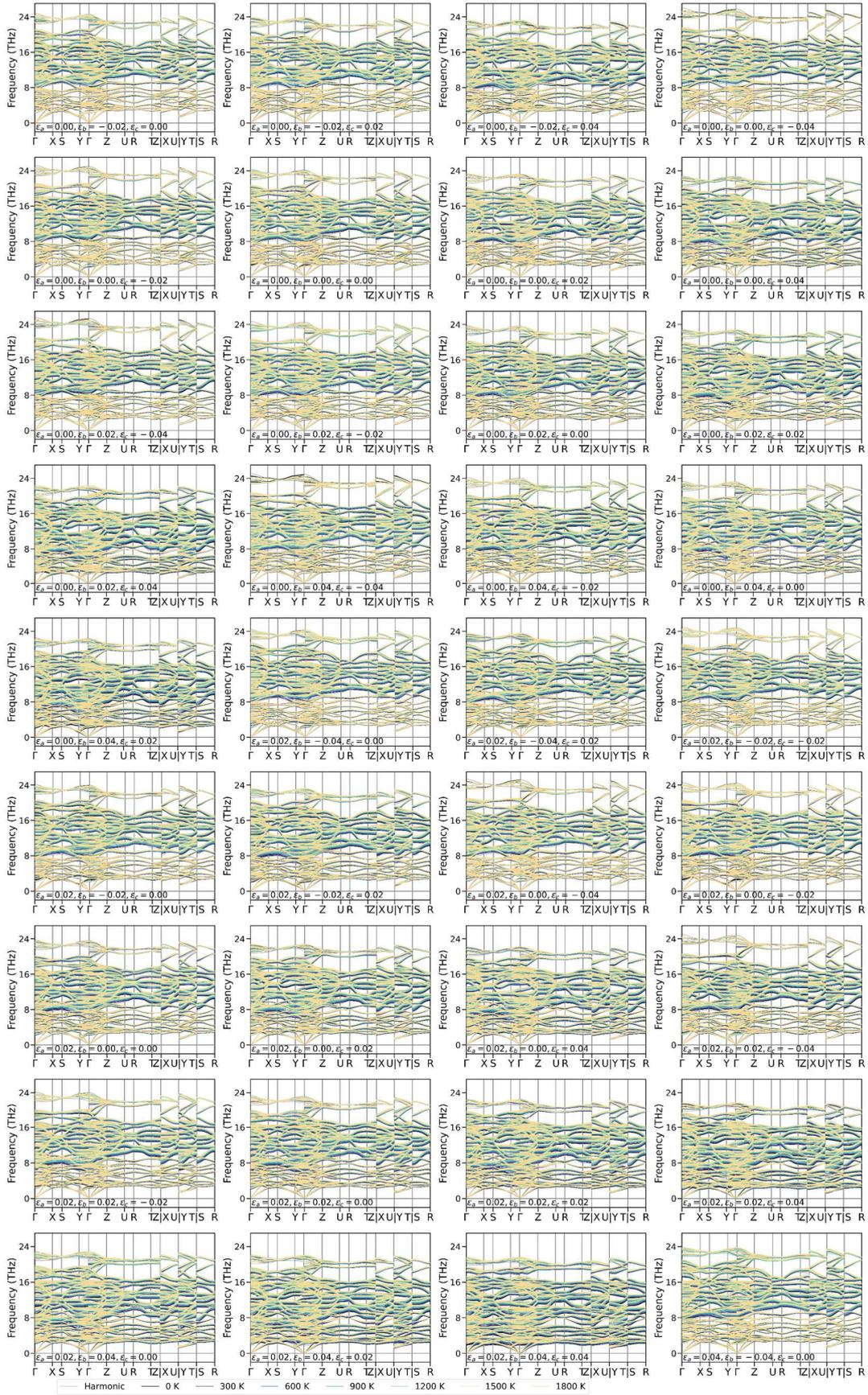

**Figure S9.** (continue) SCP spectra of oI*-HfO$_2$ under uniaxial strains $\varepsilon_a$, $\varepsilon_b$ and $\varepsilon_c$.

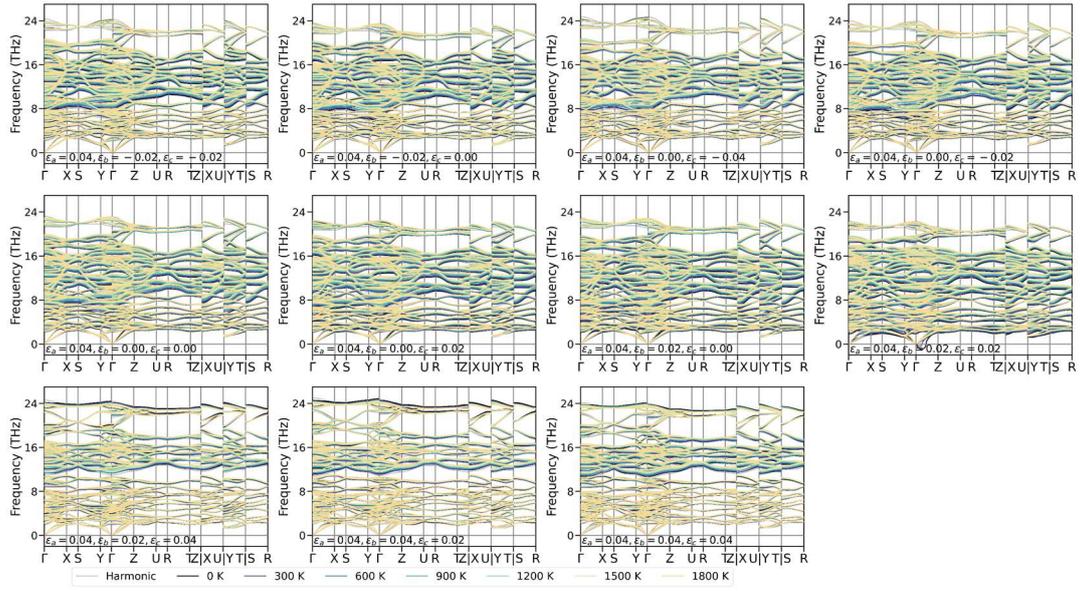

**Figure S9.** (continue) SCP spectra of oI*-HfO$_2$ under uniaxial strains $\varepsilon_a$, $\varepsilon_b$ and $\varepsilon_c$.

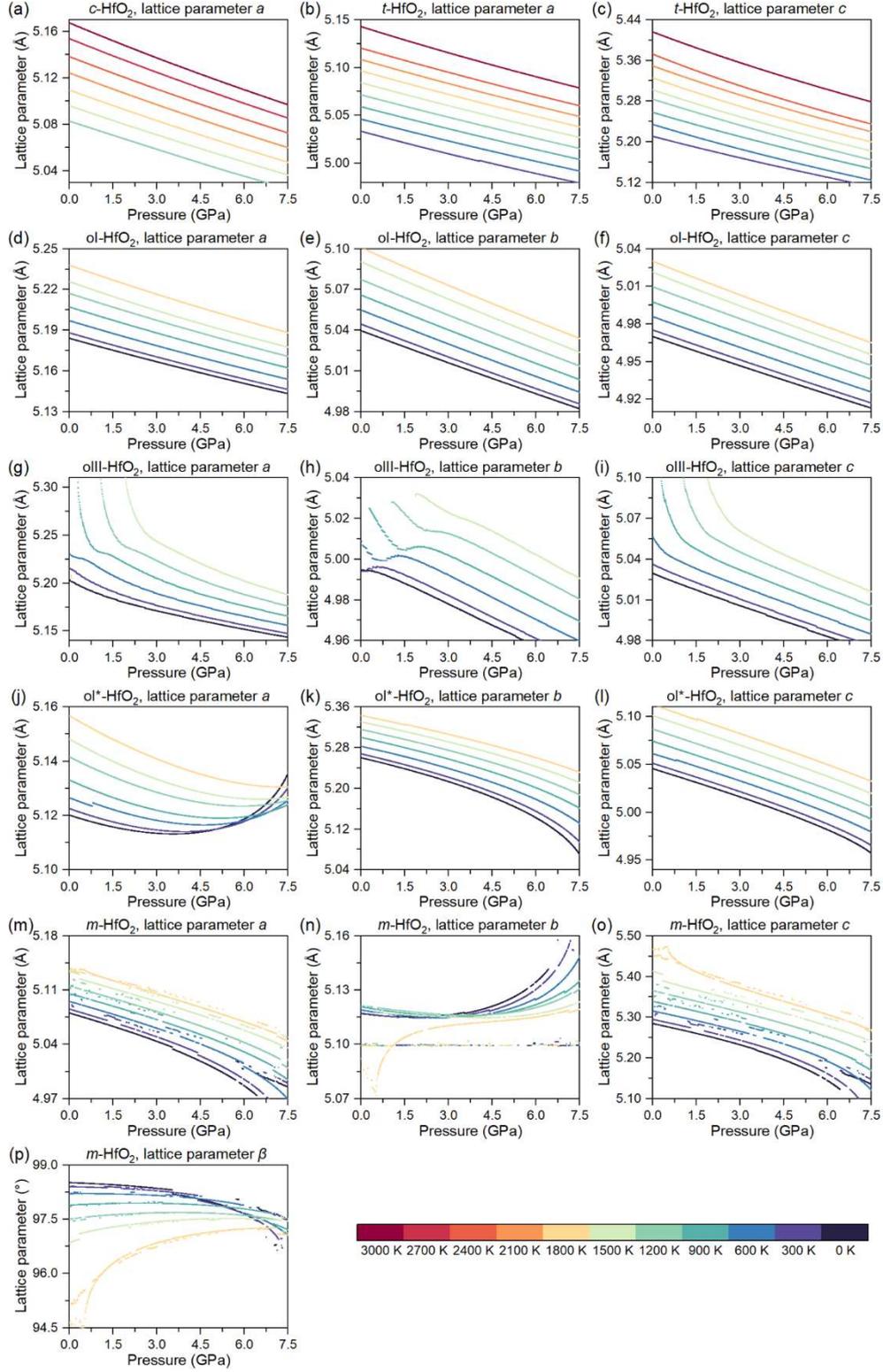

**Figure S10.** Evolution of lattice parameter (a) *a* of *c*-HfO$_2$, (b-c) *a* and *c* of *t*-HfO$_2$, (d-f) *a*, *b* and *c* of oI-HfO$_2$, (g-i) *a*, *b* and *c* of oIII-HfO$_2$ and (j-m) *a*, *b*, *c* and *β* of m-HfO$_2$ over temperature and pressure. Lattice parameters *a*, *b* and *c* of the tetragonal, orthorhombic and monoclinic phases are taken from the quasi-cubic lattice commensurate with the unit cell of *c*-HfO$_2$. Missing curve (discontinuous data) indicates diverged (ambiguous) PES minimization.